\documentclass[prl,aps,twocolumn,showpacs]{revtex4}

\usepackage{graphicx}

\begin{document}

\title{Reply to the Comment by Sandvik, Sengupta, and Campbell
on ``Ground State Phase Diagram of a Half-Filled
One-Dimensional Extended Hubbard Model''}

\author{Eric Jeckelmann} 
\affiliation{Institut f\"{u}r Physik, Johannes 
Gutenberg-Universit\"{a}t, 55099 Mainz, Germany}

\date{\today}


\maketitle

In their Comment~\cite{sandvik}, Sandvik, Sengupta, and Campbell
present some numerical evidences to support the existence of an 
extended bond-order-wave (BOW) phase at couplings $(U,V)$ 
\emph{weaker} than a 
tricritical point $(U_t,V_t)$ \cite{nakamura,sengupta}
in the ground state phase diagram of the one-dimensional half-filled 
$U$-$V$ Hubbard model.
They claim that their results do not agree with the phase diagram
proposed in my Letter~\cite{jeckelmann}, which shows a BOW phase for 
couplings \emph{stronger} than the critical point only.
However, I argue here that their results are not conclusive and do
not refute the  phase diagram described in the Letter.

First, while the parameter $U=4t$ used in the Comment is smaller than 
the tricritical coupling $U_t$ found in Ref.~\cite{sengupta}, 
it is larger than 
other estimations of $U_t$ (see references in the Letter). 
Therefore, results for $U=4t$ only are not sufficient
to determine the position of the BOW phase with respect to
the tricritical point, which is the most important qualitative
difference between the phase diagram in the Letter
and those described in Refs.~\cite{nakamura,sengupta}.
To prove the existence of a BOW phase at couplings weaker than the
tricritical point, one should use parameters $U$ 
smaller than any estimation of $U_t$.  

Second, the finite-size-scaling analysis of the charge susceptibility
$\chi_c(q)$ in Fig.~1(a) of the Comment is misleading. 
A correct analysis is to take the limit $N\rightarrow \infty$ first
and then look at the $q \rightarrow 0$ limit.
Sandvik, Sengupta, and Campbell takes both limits simultaneously
($q=2\pi/N$), which can lead to incorrect results. 
For instance, the function $F_N(q) = 1/(qN)$ vanishes
if the limit $N \rightarrow \infty$
is taken first, but tends to a 
constant $1/2\pi$ if both limits are taken simultaneously. 
Thus, the results shown in the Comment are no 
proof of a continuous phase transition as a function of $V$ for $U=4t$.

Third, although I can not rigorously exclude the existence
of an extended BOW region in the phase diagram, 
my results show that its width would certainly be much smaller than 
predicted in Ref.~\cite{nakamura}.
The main features of the BOW phase (as compared to the competing
Mott insulator phase) are (i) a long-range-ordered BOW (dimerization)
and (ii) a spin gap. 
I have found a vanishing spin gap in the thermodynamic limit
for the example presented in Fig.~1(b) of the Comment.
In their previous work~\cite{sengupta}, Sengupta, Sandvik, and Campbell
did not present any conclusive evidence for the opening of a spin gap
in an extended region outside the charge-density-wave (CDW) regime.
It is possible that the spin gap is too small to be detected
in the finite systems investigated ($N \leq 1024$ sites), but
it is as likely that finite-size effects and an arbitrary 
extrapolation to the infinite system limit are responsible
for the rather small dimerization reported in the Comment. 
I consider that the existence of the BOW phase is demonstrated
only in those cases for which numerical results are consistent.
In particular,  both the extrapolated spin gap
and the extrapolated dimerization should be clearly larger than zero.

Fourth, the discrepancies between Sandvik, Sengupta, 
and Campbell results 
and my results are certainly not a failure of the DMRG method nor an 
effect of
open boundary conditions.
In the ground state, the staggered
bond order of an open finite chain is always larger
than in a corresponding periodic system because of the Friedel 
oscillations
induced by the chain edges.
For both types of boundary conditions the staggered bond order obtained
with DMRG
decreases with increasing numerical accuracy (i.e., an increasing
number $m$ of density-matrix eigenstates kept).
Thus, DMRG results for an open finite system
systematically \emph{overestimate} the dimerization of the infinite 
system.
The likely cause of the discrepancies is the difficulty in
extrapolating numerical results to the thermodynamic limit
in the critical region $U \approx 2V$.

Finally, the most significant finding in my Letter is the presence
of the BOW phase at couplings clearly stronger than
the tricritical point.
This fundamentally contradicts the theory~\cite{nakamura}
predicting an extended BOW phase only 
at couplings weaker than $(U_t,V_t)$.
Nevertheless, Sandvik, Sengupta, and Campbell do not dispute this 
finding
nor provide any explanation for this failure of the theory that they
claim to confirm in their Comment.

In conclusion, 
none of the numerical results presented in the Comment refute 
the conclusions of my Letter.
While the phase diagram presented in the Letter
is partially based on some hypotheses, it is supported
by reliable numerical results and a consistent theory.

\end{document}